# Design and Simulation of a Highly Sensitive Refractive Index Sensor based on Grating-assisted Strip Waveguide Directional Coupler

Parvinder Kaur and M. R. Shenoy
Department of Physics, Indian Institute of Technology Delhi, New Delhi-110016, India
Corresponding author's email: parvinder.kaur@physics.iitd.ac.in

*Abstract*— A highly sensitive refractive index sensor based on grating-assisted strip waveguide directional coupler is proposed. The sensor is designed using two coupled asymmetric strip waveguides with a top-loaded grating structure in one of the waveguides. Maximum light couples from one waveguide to the other at the resonance wavelength, and the change in resonance wavelength with the change in refractive index of the medium in the cover region is a measure of the sensitivity. The proposed sensor would be an on-chip device with a high refractive index sensitivity of ~ $10^4$ nm/RIU, and negligible temperature sensitivity (< 1nm/$^0$C). The sensor configuration offers a low propagation loss, thereby enhancing the sensitivity. Variation of the sensitivity with the waveguide parameters of the proposed sensor have been studied to optimize the design.

## I. INTRODUCTION

Application of silicon photonics based devices as a refractive index sensor has attracted considerable attention because of their fast response time, compactness, high compatibility with on-chip devices, higher sensitivity and stability. Some of these silicon photonic sensors employ directional couplers [1], coupled slot waveguides [2], and micro-ring resonators [3]. Such refractive index sensors find applications in bio-sensing and chemical sensing [4].

Generally, there are two optical methods used for sensing: the label based sensing and the label-free sensing [5]. Label based sensing requires the sample preparation before the sensing, whereas no sample preparation is required in case of label-free sensing. The label-free guided wave optical sensors are based on interaction with the evanescent field present in the cladding of the optical waveguide. If the sensing material is impressed in the cladding (or if it forms the cladding) of the waveguide, the propagation characteristics of the waveguide are directly influenced by the perturbation in the sensing area. Some of the sensors which use the label-free method are refractive index sensor [6], absorption sensor [7] and temperature sensor [8]. Various waveguide designs such as slot, strip and rib waveguides in silicon photonics have been proposed to increase the sensitivity of the sensor [9-10]. Till date, slot waveguide structures are extensively used for silicon photonics based refractive index sensors [11-14].

Grating-assisted strip waveguide directional coupler based components have found various applications, such as contra directional coupler [15], micro ring resonator [16], optical wavelength filter [17], etc. Grating-assisted slot-strip waveguide structure has also been studied for sensor applications [18]. The sensitivity of the sensor is also affected by the propagation loss, which broadens the peaks at the resonance wavelength, which in turn reduces the sensitivity of the sensor. These losses are usually due to the sidewall roughness and leakage into the substrate. Since the propagation losses are higher in slot waveguides, in comparison to that in strips waveguides [19], the sensitivity of strip waveguide based sensors are expected to be higher.

In this paper, we propose a highly sensitive refractive index sensor based on grating-assisted strip waveguide directional coupler. We have studied the refractive index sensitivity and temperature dependence of the sensor by tailoring the waveguide parameters, and found that the proposed configuration offers high refractive index sensitivity and very low temperature sensitivity. We have studied the propagation losses of the strip waveguide and optimized the width for low propagation loss. In Section II, we have presented the configuration and principle of working of the proposed sensor. Section III includes simulation results and discussion. Section IV describes the temperature dependence of the sensor, followed by the propagation losses in Section V. The findings are summarized in conclusion.

## II. Sensor Configuration and Working Principle

The proposed waveguide sensor (see Fig. 1) comprises of two parallel non-identical strip waveguides of $Si_3N_4$, over $SiO_2$ substrate, placed side by side with a separation of 2*d*. A refractive index grating is formed on top of one of the waveguides of the directional coupler, and with an overlay of $SiO_2$. The period of the grating is chosen such that the phase-shift introduced by the grating provides phase matching between the two non-synchronous waveguides. The two waveguides have different effective refractive indices (i.e. non-synchronous in phase) of the two guided modes, and very little evanescent coupling would take place between them. Therefore, the grating is used to achieve phase matching between these two modes. The grating period $Λ$ is chosen, such that maximum light gets transferred from one waveguide to the other at a wavelength $λ_r$, called the resonance wavelength; the corresponding phase matching condition is given by

$$λ_r = (n_{eff1} - n_{eff2})Λ \quad (1)$$

where $n_{eff1}$ and $n_{eff2}$ are the effective indices of the two normal modes of the coupled waveguide structure. Eq. (1) shows that the grating-assisted directional coupler is wavelength selective. If a broad-band light is launched into Waveguide 1, light at the resonance wavelength $λ_r$ gets coupled into Waveguide 2, with a band-pass spectrum, and the residual power, corresponding to the band-rejection spectrum remains in Waveguide 1. As the refractive index of the external medium changes, the effective indices of the two modes change and the resonance wavelength shifts. The sensitivity of the sensor is given by [20]:

$$\frac{dλ_r}{dn_{ex}} = λ_r \frac{ΔN}{ΔN_g} \quad \text{with} \quad ΔN = \frac{\partial n_{eff1}}{\partial n_{ex}} - \frac{\partial n_{eff2}}{\partial n_{ex}} \quad (2)$$

$$ΔN_g = N_{g1} - N_{g2} \quad (3)$$

$$N_{g1} = n_{eff1} - λ\left(\frac{dn_{eff1}}{dλ}\right); N_{g2} = n_{eff2} - λ\left(\frac{dn_{eff2}}{dλ}\right) \quad (4)$$

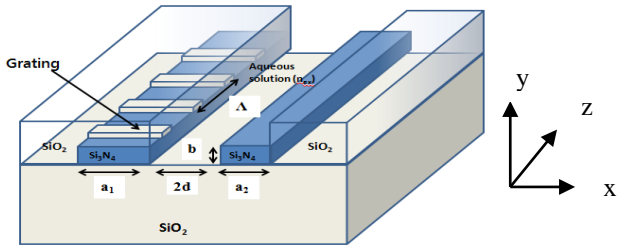

*Fig.1: Schematic of the proposed refractive index sensor. $a_1$ and $a_2$ are the widths of the strip Waveguides 1 and 2, respectively; 2d represents the spacing between the two waveguides; b is the height of the waveguides; and Λ is the period of the grating.*

$N_{g1}$ and $N_{g2}$ are the *group indices* of the two normal modes; $ΔN$ represents the difference in the variation of the effective indices of the two modes with the refractive index of the external medium.

## III. Simulation Results and Discussion

Figures 2 (a) and (b) show the electric field ($E_x$) distributions in the two waveguides which are numerically computed by using full vectorial approach (COMSOL). The coupled waveguide structure supports both TE (Transverse Electric) and

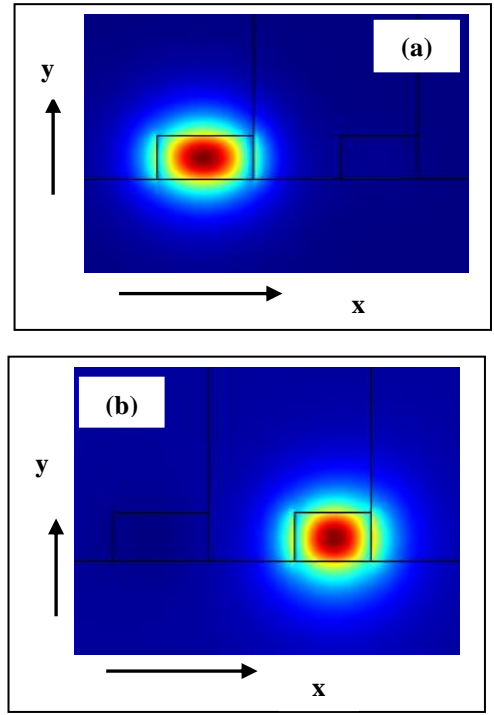

*Fig. 2: Electric field ($E_x$) distributions of the normal modes across the two strip waveguides of the directional coupler.*

TM (Transverse Magnetic) polarizations, and we have considered both polarizations in the following sensor analysis. The waveguide parameters used in the sensor configurations are $a_1 = 1\mu m$, $a_2 = 800$nm, $2d = 900$nm, $b = 455$nm, $n_{ex} = 1.333$ and resonance wavelength $\lambda_r = 1.55\mu m$: these parameters are typical of single mode $Si_3N_4$ waveguides fabricated on $SiO_2$ substrate [10]. Figure 3 shows the variation of the sensitivity with the height $b$ of the waveguides. As expected the sensitivity decreases with increasing values of $b$. We have observed that the sensitivity remains high with a good confinement of the field for ~ 455nm. With these parameters, the sensitivity of the sensor for TE and TM polarizations are found to be $1.59\times10^4$nm/RIU and $1.11\times10^4$nm/RIU, respectively, which is almost 10 times higher than that of the sensitivity reported in ref [18].

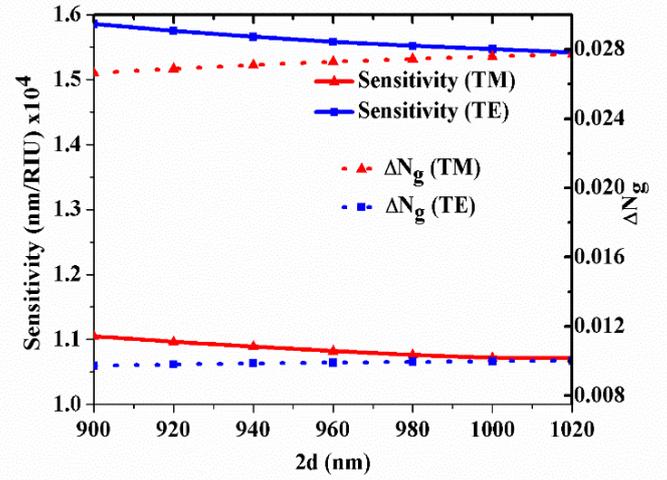

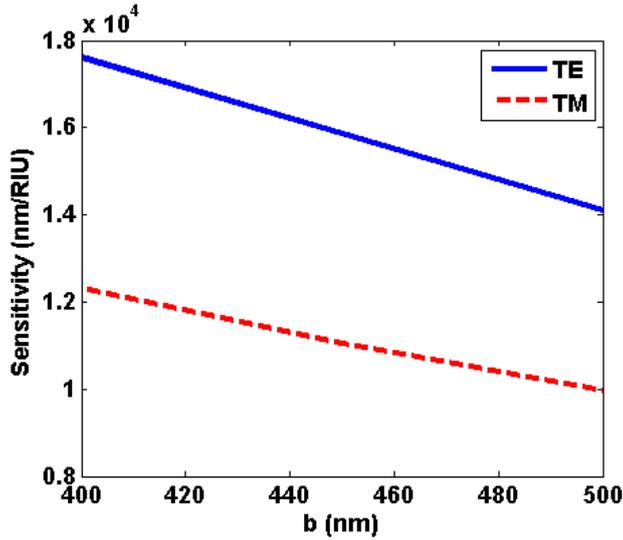

Fig. 3: Variation of the sensitivity with the height b of the waveguide.

We further optimized the spacing between the two waveguides and width of the Waveguide 2. Figure 4 shows the variation of the group index difference and the corresponding variation of the sensitivity with the spacing between the two waveguides. The group index difference of the TE polarization is lower than that for the TM polarization, and it slightly increases with the spacing between the two waveguides. Since the sensitivity is inversely proportional to the group index difference ($\Delta N_g$), the sensitivity decreases slightly with the spacing between the two waveguides, and it is slightly higher for the TE polarization than that for the TM polarization. Figure 5 shows the variation of the group index difference with the width of Waveguide 2 for both TE and TM polarizations. The group index difference increases with the width $a_2$ of Waveguide 2. It changes sign from negative to positive as $a_2$ increases, and it is close to zero when $a_2$ is ~ 855nm and ~ 935nm for TE and TM polarizations, respectively.

Fig. 4: (right) Variation of the group index difference and (left) sensitivity with the spacing (2d) between the two waveguides for TE and TM polarizations.

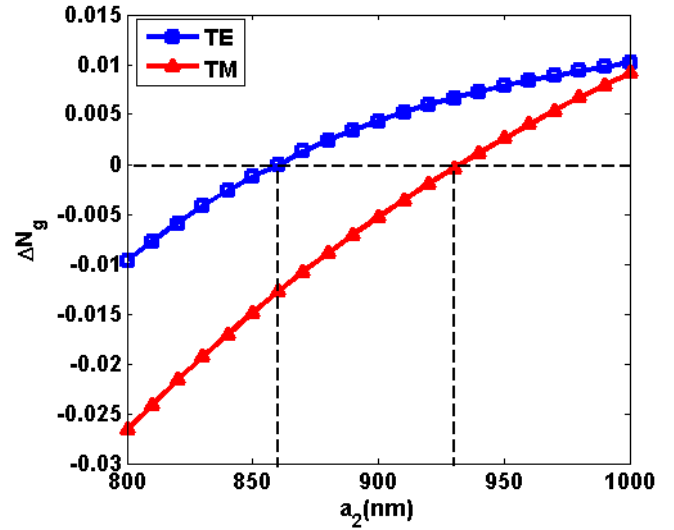

Fig. 5 Variation of the group index difference ($\Delta N_g$) with the width $a_2$ of the Waveguide 2.

Figure 6 shows the variation of the refractive index sensitivity with the width $a_2$ of Waveguide 2 for TE (a) and TM (b) polarizations. Since the sensitivity is inversely proportional to the group index difference ($\Delta N_g$), it increases drastically when the group index difference is close to zero. Hence one can achieve very high sensitivity just by tuning the width $a_2$ of Waveguide 2. Figure 7 shows the dynamic range of the sensor, for three values of $a_2$ = 800nm, 850nm and 900nm. The sensitivity increases with the refractive index of the external medium and it peaks at refractive index values of 1.60, 1.55 and 1.50 for waveguide widths ($a_2$) 800nm, 850nm and 900nm, respectively. As we further increase the refractive index of the external medium, the sensitivity starts decreasing because of severe degradation of mode confinement in the core of Waveguide 2, which limits the dynamic range.

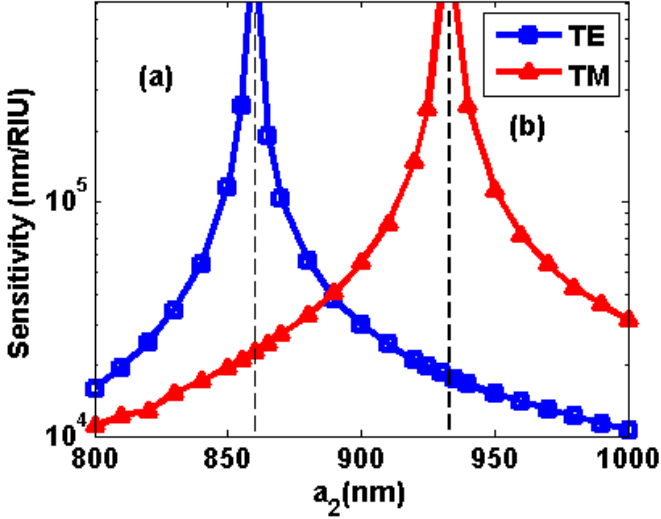

*Fig. 6: variation of the sensitivity with the width of Waveguide 2, (a) TE polarization and (b) TM polarization.*

**Fig. 6: variation of the sensitivity with the width of Waveguide 2, (a) TE polarization and (b) TM polarization.**

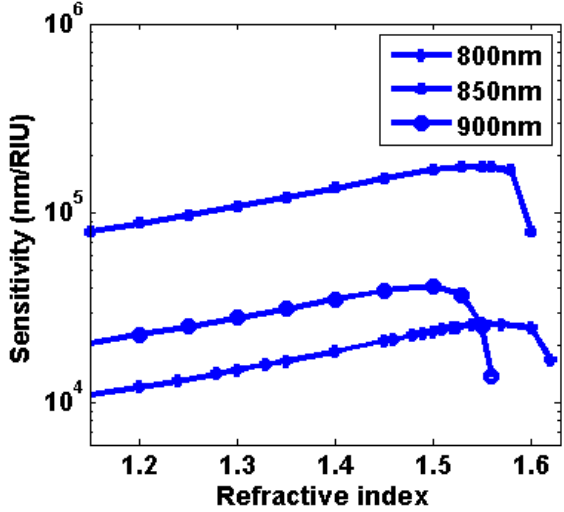

*Fig. 7: Variation of the sensitivity with the refractive index of the external material.*

## IV. Temperature Sensitivity:

The temperature sensitivity of the resonance wavelength is proportional to the temperature dependence of the effective refractive indices, and can be derived from Eq. (1), and expressed as [21]

$$\frac{d\lambda_r}{dT} = \Lambda\left(\frac{dn_{eff1}}{dT} - \frac{dn_{eff2}}{dT}\right) + (n_{eff1} - n_{eff2})\frac{d\Lambda}{dT} \quad (5)$$

where $T$ is the temperature. The second term in the RHS of Eq. (5) can be neglected because the shift in the resonance wavelength is very small. The refractive indices of the core, cladding and external medium at a particular temperature $T$ can be written as $n_{co}(T) = n_{co}(T_0) + \varepsilon_{co}\Delta T$, $n_{cl}(T) = n_{cl}(T_0) + \varepsilon_{cl}\Delta T$ and $n_{ex}(T) = n_{ex}(T_0) + \varepsilon_{ex}\Delta T$ respectively, where $\Delta T = T - T_0$ and $\varepsilon_{co} = 4 \times 10^{-5}/^0C$, $\varepsilon_{cl} = 1 \times 10^{-5}/^0C$ and $\varepsilon_{ex} = -8 \times 10^{-5}/^0C$ are the thermo-optic coefficients of the core, cladding and the external medium, respectively [18]. Considering the wavelength dependence of the effective indices, Eq. (5) becomes:

$$\frac{d\lambda_r}{dT} = \frac{\lambda_r}{\Delta N_g}\left(\frac{\partial \Delta n_{eff}}{\partial n_{co}}\varepsilon_{co} + \frac{\partial \Delta n_{eff}}{\partial n_{cl}}\varepsilon_{cl} + \frac{\partial \Delta n_{eff}}{\partial n_{ex}}\varepsilon_{ex}\right) \quad (6)$$

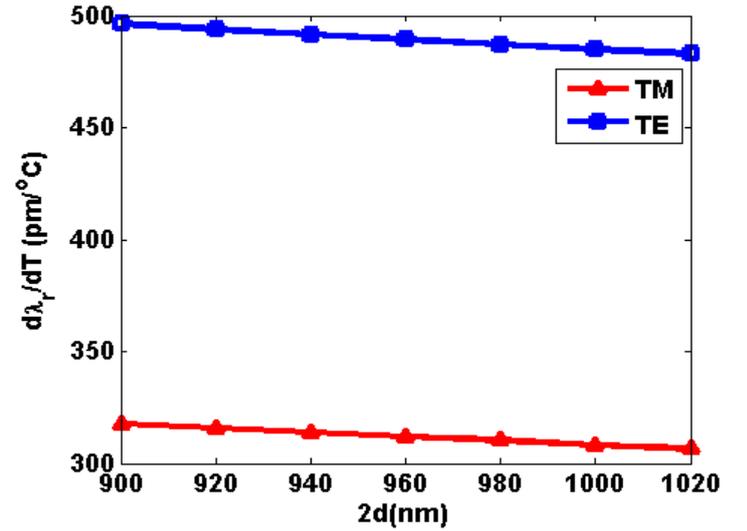

*Fig. 8: Variation of the temperature sensitivity of the resonance wavelength with the spacing between the two waveguides of the directional coupler for TE and TM polarizations.*

Figure 8 shows the variation of the temperature sensitivity with the spacing between the two strip waveguides; all parameters used are the same as those used in Fig. 4. For both TE and TM polarizations, it decreases slightly as we increase the spacing between the two waveguides

## V. Propagation loss

To study the propagation loss of the proposed sensor, we have used a numerical model in which the experimentally measured propagation loss for a set of parameters are considered as a reference to estimate the propagation loss for the proposed waveguide structure. It was introduced for the first time in the ref [22] and shown to be in good agreement with the measured values. As a reference, we have used the measured propagation loss of 4dB/cm for a $Si_3N_4$ waveguide of width 1µm and height 500nm with $SiO_2$ cladding [23]. In the absence of experimental data on propagation loss in $Si_3N_4$ waveguides for TE and TM modes, we have assumed about 25% higher loss for the TM modes to estimate the propagation loss of the proposed structure. This is reasonable considering the reported results on the differential loss in TE and TM modes in strip waveguides of other materials [22-24]. In the waveguide structure analysed by COMSOL, two side walls of width $t = 10$nm are introduced on each side of the waveguide (see Fig. 9). The propagation loss due to scattering from the side walls of the waveguide are included by adding an imaginary part of the refractive index for the wall region in the model. The propagation loss in dB can be write in the form [22]

$$\alpha = \frac{54.57 n_{eff\_img} z}{\lambda_r} \quad (7)$$

where $n_{eff\_img}$ is imaginary part of the effective index and $z$ is the length of the waveguide.

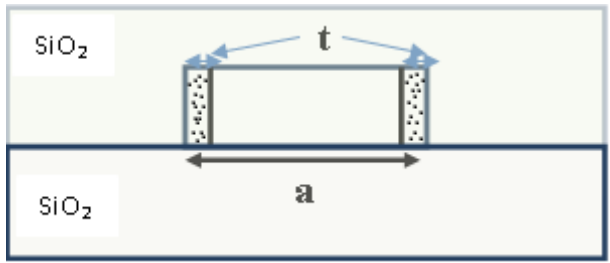

*Fig. 9: Schematic of the cross section of the waveguide structure, used in simulation.*

First, the simulation was done in COMSOL to iterate the complex refractive index of the side walls corresponding to a measured value of the propagation loss. Once we get the complex refractive index ($n_{core}+in_{wall}$, where $n_{core}$ is the refractive index of the core and $n_{wall}$ is the imaginary part of the refractive index, representing the loss due to wall roughness) of the wall region. Using the complex refractive index of the waveguide, we have calculated the complex effective index of the propagating mode, $n_{eff\_r} - in_{eff\_img}$. The imaginary part of the effective index is then used to estimate the propagation loss for the proposed waveguide structure. Figure 10 shows the variation of the propagation losses with the width of the waveguide. The propagation loss decreases as we increase the width of the waveguide, for a fixed wall thickness. To reduce the propagation loss, one can increase the width of the waveguide, however it should remain single moded. Figure 11 shows the variation of the propagation loss with the wall thickness. As expected, the propagation loss increases with the wall thickness. However, it was tested that the simulation results are independent of the wall thickness $t$ by carrying out the simulation for the wall thickness $t$ = 5nm and 10nm. In our calculation, we have used the wall thickness 10nm.

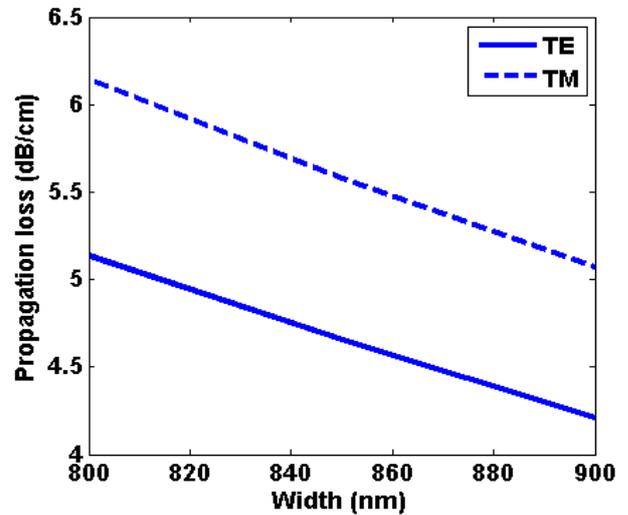

*Fig. 10: Variation of the propagation losses with the width of the waveguide for both TE and TM polarization.*

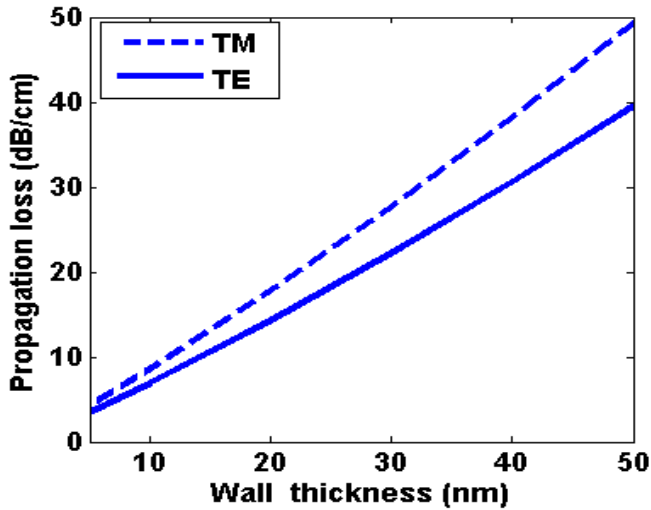

*Fig. 11: Variation of the propagation loss with the wall thickness of the waveguide.*

## VI. CONCLUSION

We have presented the design and simulation results of a refractive index sensor based on grating-assisted strip waveguide directional coupler. The sensor consists of two coupled asymmetric strip waveguides, and the phase matching is achieved by the grating structure. The proposed sensor is highly sensitive with a sensitivity of ~ $10^4$ nm/RIU. The dynamic range of the sensor has also been estimated. We have also studied the variation of the sensitivity with the waveguide parameters and the temperature dependence of the sensor. The propagation losses of the strip waveguides are minimized to achieve a better sensitivity.

**References:**
1. P. Dumais, C. L. Callender, J. P. Noad, and C. J. Ledderhof, "Silica-on-Silicon Optical Sensor Based on Integrated Waveguides and Micro-channels", IEEE Photonics Technology Letters, **17**, pp. 441-443, (2005).
2. S. Nacer and A. Aissat, "Optical sensing by silicon slot-based directional couplers", Optical and Quantum Electronics, **44**, pp. 35-43, (2012).
3. K. De Vos, I. Bartolozzi, E. Schacht, P. Bienstman, and R. Baets, "Silicon-on-insulator micro-ring resonator for sensitive and label-free biosensing", Optics Express **15**, pp. 7610-7615, (2007).
4. J. Vorosa, J. J. Ramsden, G. Csucs, I. Szendro, S. M. De Paul, M. Textor, N. D. Spencer, "Optical grating coupler biosensors", Biomaterials, **23**, pp. 3699–3710, (2002)
5. A. Syahir, K. Usui, K. Tomizaki, K. Kajikawa and H. Mihara, "Label and Label-Free Detection Techniques for Protein Microarrays", Microarrays, **4**, pp. 228-244, (2015).
6. X. D. Fan, I. M. White, S. I. shopova, H. Y. Zhu, J. D. Suter, and Y. Z. Sun, "Sensitivity optical biosensors for unlabeled targets a review" Anal. Chim. Acta, **620**, pp. 8-26 (2008).
7. A. Og. Dikovska, G.B. Atanasova, N.N. Nedyalkov, P.K. Stefanov, P.A. Atanasov, E.I. Karakolevac, A.Ts. Andreev, "Optical sensing of ammonia using ZnO nanostructure grown on a side-polished optical-fiber", Sensors and Actuators B, 146, pp. 331–336, (2010).
8. Qing Liu, K. S. Chiang, K. P. Lor, and C. K. Chow, "Temperature sensitivity of a long-period waveguide grating in a channel waveguide", Applied Physics Letters, **86**, pp. 241115 (2005).
9. S.M. Topliss, S.W. James, F. Davis, S.P.J. Higson, R.P. Tatam, "Optical fibre long period grating based selective vapour sensing of volatile organic compounds", Sensors and Actuators B, **143**, pp. 629–634, (2010).
10. Q. Liua, X. Tua, K. W. Kima, J. S. Keea, Y. Shina, K. Hana, Y. J. Yoonb, G. Q. Loa, M. K. Park, "Highly sensitive Mach–Zehnder interferometer biosensor based on silicon nitride slot waveguide", Sensors and Actuators B, **188**, pp. 681–688, (2013).
11. J. Witzens, T. B. Jones, and M. Hochberg, "Design of transmission line driven slot waveguide Mach-Zehnder interferometers and application to analog optical links", OPTICS EXPRESS, **18**, pp. 16902-16928, (2010).
12. C. A. Barrios, "Optical Slot-Waveguide Based Biochemical Sensors", sensors, **9**, pp.4751-4765, (2009).
13. T. Baehr-Jones and M. Hochberg, "Optical modulation and detection in slotted Silicon waveguides", Optics Express, **13**, pp. 5216-5216, (2005).
14. V. M. N. Passaro, F. D. Olio, C. Ciminelli and M. N. Armenise, "Efficient Chemical Sensing by Coupled Slot SOI Waveguides", Sensors, **9**, pp. 1012-1032, (2009).
15. Wei Shi, Xu Wang, Charlie Lin, Han Yun, Yang Liu, Tom Baehr-Jones, Michael Hochberg, Nicolas A. F. Jaeger, and Lukas Chrostowski, "Silicon photonic grating-assisted, contra-directional couplers", Optics Express, **22**, pp. 3633-3650, (2013).
16. W. Shi, X. Wang, W. Zhang, H. Yun, C. Lin, L. Chrostowski, and N. A. F. Jaeger, "Grating-coupled silicon microring resonators", Applied Physics Letters, **100**, pp. 121118, (2012).

17. Daoxin Dai and John E. Bowers, "Silicon-based on-chip multiplexing technologies and devices for Peta-bit optical interconnects", Nanophotonics, 3, pp. 283–311, (2014).
18. Q. Liu, J. Sheng Kee, and M. K. Park, "A refractive index sensor design based on grating-assisted coupling between a strip waveguide and a slot waveguide", Optics Express, **21**, pp. 5897-5909 (2012).
19. R. Palmer, L. Alloatti, D. Korn, W. Heni, P. C. Schindler, J. Bolten, M. Karl, M. Waldow, T. Wahlbrink, W. Freude, C. Koos, J. Leuthold, "Low-Loss Silicon Strip-to-Slot Mode Converters", IEEE Photonics Journal, **5**, pp. 2200409-2200509, (2013).
20. X. Sui, L. Zhang, and I. Bennion, "Sensitivity characteristics of long-period fiber gratings", Journal of Lightwave Technology, **20**, pp. 255-266, (2002).
21. M. N. Ng, K. S. Chiang, "Thermal effects on the transmission spectra of long-period fiber gratings", Optics Communications, **208**, pp. 321-327, (2002).
22. S. M. Lindecrantz, and O. G. Helleso, "Estimation of Propagation Losses for Narrow Strip and Rib Waveguides", IEEE Photonics Technology Letters, **26**, pp. 1836-1839, (2014).
23. D. T. H. Tan, K. Ikeda, P. C. Sun, and Y. Fainman, "Group velocity dispersion and self-phase modulation in silicon nitride waveguides", Applied Physics Letters, **96**, pp. 61101-61103, (2010).
24. N. Daldosso, M. Melchiorri, F. Riboli, M. Girardini, G. Pucker, M. Crivellari, P. Bellutti, A. Lui, and L. Pavesi, "Comparison Among Various $Si_3N_4$ Waveguide Geometries Grown Within a CMOS Fabrication Pilot Line", J. Lightwave Technology, **22**, pp. 1734 - 1740, (2004)